\documentclass[proof]{pasj01}
\draft % <<<<<<<<<<<<<<<<<<<<<<<<<

\usepackage{graphicx}
\begin{document} 
\Received{2017/06/16}%{yyyy/mm/dd}
\Accepted{}%{yyyy/mm/dd}
%\Published{yyyy/mm/dd}

\title{Bipolar H{\sc ii} regions produced by cloud/cloud collisions}

%%% begin:list of authors
\author{Anthony \textsc{Whitworth}\altaffilmark{1}}
\altaffiltext{1}{School of Physics \& Astronomy, Cardiff University, CF24 3AA, Wales, UK}
\email{ant@astro.cf.ac.uk}
\author{Oliver \textsc{Lomax}\altaffilmark{1}}
\author{Scott \textsc{Balfour}\altaffilmark{1}}
\author{Pierre \textsc{M{\`e}ge}\altaffilmark{2}}
\altaffiltext{2}{Aix Marseille Univ, CNRS, LAM, Laboratoire d'Astrophysique de Marseille, Marseille, France}
\author{Annie \textsc{Zavagno}\altaffilmark{2}}
\author{Lise \textsc{Deharveng}\altaffilmark{2}}

%% `\KeyWords{}' always has to be placed before `\maketitle'.
 %Do NOT move this preamble from here!

\maketitle

%%%%%
\begin{abstract}
We suggest that Bipolar H{\sc ii} Regions may be the aftermath of collisions between clouds. Such a collision will produce a shock-compressed layer, and a star cluster can then condense out of the dense gas near the centre of the layer. If the clouds are sufficiently massive, the star cluster is likely to contain at least one massive star, which emits ionising radiation, and excites an H{\sc ii} region, which then expands, sweeping up the surrounding neutral gas. Once most of the matter in the clouds has accreted onto the layer, expansion of the H{\sc ii} Region meets little resistance in directions perpendicular to the mid-plane of the layer, and so it expands rapidly to produce two lobes of ionised gas, one on each side of the layer. Conversely, in directions parallel to the mid-plane of the layer, expansion of the H{\sc ii} Region stalls due to the ram-pressure of the gas that continues to fall towards the star cluster from the outer parts of the layer; a ring of dense neutral gas builds up around the waist of the Bipolar H{\sc ii} Region, and may spawn a second generation of star formation. In this paper we present a dimensionless model for the flow of ionised gas in a Bipolar H{\sc ii} Region created according to the above scenario, and predict the characteristics of the resulting freefree continuum and recombination-line emission. This dimensionless model can be scaled to the physical parameters of any particular system. Our intention is that these predictions will be useful in testing the scenario outlined above, and thereby providing indirect support for the role of cloud/cloud collisions in triggering star formation.
\end{abstract}
%%%%%

%%%%%
\section{Introduction}
%%%%%

It has long been suspected that collisions between clouds and/or turbulent streams play a critical role in triggering star formation. (We do not wish to distinguish here between a cloud/cloud collision and a collision between two turbulent streams; all that we have in mind is an interaction between two anti-parallel flows that delivers interstellar matter into a local high-density state, where its self-gravity becomes important and leads to star formation.)

This picture is intrinsic to the perception that spiral arms are lit up by newly-formed massive stars, because orbit-crowding in the arm increases the chance of cloud/cloud collisions there, and it is intrinsic to many other models that seek to explain why star formation is concentrated in specific locations rather than being distributed throughout the disc of the Milky Way (e.g. \cite{StoneM1970a, StoneM1970b, ShuFetal1972, LorenRB1976, RobeHaus1984, HausRobe1984, KwanVald1983, KwanVald1987, Chapetal1992, Pongetal1992, HabeOhta1992, Whitetal1994a, Whitetal1994b, Turnetal1995, Whitetal1995, Bhatetal1998, Balletal1999, Heitetal2005, Balletal2006, Heitetal2006, Vazqetal2007, Balletal2007, Heitetal2008, Torietal2011, Dobbetal2014, Fukuetal2014, Dobbetal2015, Hawoetal2015a, Hawoetal2015b, Fukuetal2015, Balfetal2015, Fukuetal2016, Torietal2017, Balfetal2017})

It is also intrinsic to the notion of turbulent fragmentation, whereby, within a molecular cloud, the locations of star formation are the places where intermittently the chaotic motions deliver parcels of gas, either directly into a gravitationally unstable pre-stellar core, or into a filament that will then guide the gas into such a core (e.g. \cite{Klesetal2000, Heitetal2001, KlesBurk2001, KlessenR2001, PadoNord2002, Bateetal2003, PadoNord2004, Claretal2011, HennChab2011, PadoNord2011, FedeKles2012, HennChab2013, Bertetal2016, Whitwort2016}).

However, robust observational evidence for star formation triggered by collisions between clouds or turbulent streams is hard to obtain. If cloud masses, sizes and internal velocity dispersions subscribe to Larson's relations ($R_{_{\rm CLOUD}}\sim 0.1\,{\rm pc}\,(M_{_{\rm CLOUD}}/{\rm M}_{_\odot})^{1/2}$, $\sigma_{_{\rm CLOUD}}\sim 0.2\,{\rm km}\,{\rm s}^{-1}\,(M_{_{\rm CLOUD}}/{\rm M}_{_\odot})^{1/4}$), the  duration of a collision between two clouds with mass $M_{_{\rm CLOUD}}$ is
\begin{eqnarray}\nonumber\label{EQN:t_COLL}
t_{_{\rm COLL}}&\sim&\frac{R_{_{\rm CLOUD}}}{\sigma_{_{\rm CLOUD}}(10\,M_{_{\rm CLOUD}})}\\\label{EQN:t_COLL}
&\sim&\;0.3\,{\rm Myr}\,\left(\frac{M_{_{\rm CLOUD}}}{{\rm M}_{_\odot}}\right)^{\!1/4}\,,
\end{eqnarray}
where for simplicity we have assumed that the mean bulk velocity of a cloud is equal to the internal velocity dispersion of clouds that are ten times more massive. The time required to accumulate a gravitationally unstable layer {\it and} form stars is 
\begin{eqnarray}\nonumber
t_{_{\rm FRAG}}&\sim&2\,\left(\frac{a_{_{\rm O}}}{G\,\rho(M_{_{\rm CLOUD}})\,\sigma(10M_{_{\rm CLOUD}})}\right)^{1/2}\\\label{EQN:t_FRAG}
&\sim&\;1.4\,{\rm Myr}\,\left(\frac{M_{_{\rm CLOUD}}}{{\rm M}_{_\odot}}\right)^{\!1/8}\,,
\end{eqnarray}
Here we have assumed that the gas in the shock-compressed layer quickly cools to $T\simeq 10\,{\rm K}$, and hence has an isothermal sound speed $a_{_{\rm O}}\sim 0.2\,{\rm km}\,{\rm s}^{-1}$. \citet{Whitwort2016} shows that this is a good assumption, due to CO line cooling in the post-shock gas.

Although the values in Eqns. (\ref{EQN:t_COLL}) and (\ref{EQN:t_FRAG}) are only indicative, it follows that for all but the highest-mass clouds, the collision will often be over by the time that any stars form, so it will be hard to confirm a causal link between the collision and star formation. However, if the clouds are still colliding when stars have formed, observational confirmation requires us (i) to find two clouds which have significant overlap on the sky, and have significantly different radial velocities, (ii) to infer that the blue-shifted cloud is behind the redshifted one, (iii) to look for kinematic evidence for the gas in the layer at intermediate velocities (bridging gas), and finally, (iv) to establish that at least some of the stars in this direction would not have formed without the collision. This approach has been pioneered by \citet{LorenRB1976}, and taken up more recently by \citet{Torietal2011, Fukuetal2014, Hawoetal2015a, Hawoetal2015b, Fukuetal2015, Fukuetal2016, Torietal2017}. It may be possible to detect the newly-shocked gas in the higher-$J$ lines of molecules like CO, because these lines play an important role in cooling the post-shock gas.

A complementary approach is to look for the aftermath of a star-forming collision. \citet{Balfetal2015, Balfetal2017} and Balfour et al. (in prep.) have shown that low-velocity collisions between massive clouds tend to lead to the formation of monolithic star clusters that contain massive stars. The cluster is located near the centre of the shock-compressed layer, and once a star forms that is sufficiently massive to ionise the remaining gas in the cluster, the rate of star formation falls abruptly. The development of the resulting H{\sc ii} region is strongly influenced by the surrounding gas distribution, i.e. the shock compressed layer, and this is a direct consequence of the cloud/cloud collision.

If the ionising star remains close to the `mid-plane' of the layer (i.e. the contact discontinuity between the gas from the two clouds), or if there is more than one ionising star and at least one ends up on either side of the mid-plane, the H{\sc ii} region develops a bipolar morphology, with the two lobes on either side of the mid-plane where the density is low and the advance of the ionisation front is rapid. Conversely, in the mid-plane of the layer, the advance of the ionisation front is slow due to the high density where it meets gas falling in from the outer parts of the layer. In fact this build-up of gas at the waist of the Bipolar H{\sc ii} region may subsequently lead to a second generation of star formation. Such an H{\sc ii} region will only appear bipolar if viewed from directions close to the mid-plane of the layer; if viewed from directions far from the mid-plane, it will appear like an approximately circular shell, but the brightening at the edge will be due to the fact that the strongest emission is concentrated where the ionisation front meets the waist (and not due to limb-brightening of an essentially spherical shell).

If the ionising star or stars all end up on one side of the mid-plane, the H{\sc ii} region will be limited to that side of the layer and will be approximately circular from all angles; the only major effect of the layer will be to make the H{\sc ii} region brighter on the side where the ionised gas is being boiled off the layer, than the opposite side, and even this will only be visible from certain viewing angles.

Bipolar H{\sc ii} regions have been studied in detail by \citet{Dehaetal2015}, and clearly show the two lobes separated by a waist, where neutral gas from the layer appears to be piling up and fragmenting to form a second generation of stars, and where the emission measure due to the dense ionised gas boiling off the inside of the waist is greatest.

Cylindrical H{\sc ii} regions, which might be cases where the observer is far from the midplane and therefore sees a ring of bright emission from the waist, have been reported by \citet{BeauWill2010}.

More spherical H{\sc ii} bubbles, some of which may be cases where the ionising stars are on one side of the layer, are commonly observed (e.g. \cite{Churetal2006, Churetal2007, Simpetal2012}).

If the observed morphologies of bipolar and cylindrical H{\sc ii} regions are indeed due to massive stars located in dense layers, it is tempting to conclude that these dense layers were produced by cloud/cloud collision, and that the same collisions triggered the formation of the massive stars (and, presumably, their lower mass siblings too). They constitute a `smoking gun', pointing to a recent cloud/cloud collision.

In the present paper we pursue this idea further by building a toy model for the flow of ionised gas in a Bipolar H{\sc ii} region formed in this way, and generating maps of the resulting freefree emission, along with recombination-line profiles from different characteristic sight-lines through the region and maps of the mean, standard deviation, skewness and kurtosis of recombination lines, with a view to identifying discriminating observational signatures.

In Section \ref{SEC:Model} we introduce the parameters describing the model. In Section \ref{SEC:Flow} we derive the differential equations controlling the flow. In Section \ref{SEC:RadTrans} we describe the procedure for generating maps and line profiles. In Section \ref{SEC:Results} we present and discuss the results. In Section \ref{SEC:Conc} we summarise our main conclusions.

%%%%%
\section{Model parameters}\label{SEC:Model}
%%%%%

We define the model with reference to a 3D Cartesian frame, $\;(x,y,z),\;$ (hereafter {\it the configuration frame}) with associated unit vectors $\;(\hat{\bf i},\hat{\bf j},\hat{\bf k})$. The underlying configuration of the ionising star, the layer and the H{\sc ii} region has cylindrical symmetry about the $z$ axis, so we also define
\begin{eqnarray}
w&=&\left(x^2+y^2\right)^{1/2}\,.
\end{eqnarray}
The configuration also has reflection symmetry about the $z\!=\!0$ plane (see below).

A star is positioned at the centre of coordinates and emits ionising photons, isotropically, at a rate $\;\dot{\cal N}_{_{\rm LyC}}$.  

Except in the vicinity of the star, there is a semi-infinite plane-parallel layer of neutral gas with midplane $\;z\!=\!0,\;$ and half thickness $\;Z_{_{\rm O}},\;$ in other words, the layer is confined to $\;|z|\!<\!Z_{_{\rm O}}.\;$ For simplicity, we assume that the density in the layer is uniform, $\;\rho_{_{\rm O}},\;$ and so the surface-density of the layer is $\;\Sigma_{_{\rm O}}\!=\!2Z_{_{\rm O}}\rho_{_{\rm O}}.\;$

In the immediate vicinity of the star, there is a circular hole in the layer, where the gas is ionised and flows away towards $\;z\!=\!\pm\infty.\;$ 

The {\it Main Ionisation Front} (essentially the part that is illuminated directly by the star) is cylindrically symmetric about the $z$ axis, and -- in any plane containing the $z$ axis -- has a semi-circular cross-section, so that it is convex as seen from the star. The centre of curvature of this semi-circular cross-section is at distance $\;W_{_{\rm O}}\;$ from the star, so the {\it Main Ionisation Front} is located at
\begin{eqnarray}
\begin{array}{l}
W_{_{\rm O}}-Z_{_{\rm O}}\;\,<\;\,w\;\,<\;\,W_{_{\rm O}}\,,\\
z\;\,=\;\,\left\{Z_{_{\rm O}}^2\,-\,\left(W_{_{\rm O}}-w\right)^2\right\}^{\!1/2}\,.
\end{array}
\end{eqnarray}

The {\it Secondary Ionisation Front} (essentially the surfaces of the plane-parallel layer, where these are only illuminated by diffuse radiation) is located at
\begin{eqnarray}
\begin{array}{l}
w\;>\;W_{_{\rm O}}\,,\\
z\;=\;\pm Z_{_{\rm O}}\,.
\end{array}
\end{eqnarray}

%%%%%
\section{The flow of ionised gas}\label{SEC:Flow}
%%%%%

We assume that the number-flux of protons off the ionisation front is given by
\begin{eqnarray}\label{EQN:IF_0}
\dot{N}_{_{\rm II}}&=&n_{_{\rm II}}(5/3)^{1/2}c_{_{\rm II}}\,,
\end{eqnarray}
where $n_{_{\rm II}}$ is the density of protons {\it at the ionisation front}, and $c_{_{\rm II}}\simeq 12\,{\rm km}\,{\rm s}^{-1}$ is the isothermal sound speed in the ionised gas; basically this means that the ionisation front is always D-critical. We also assume (i) that the flow of ionised gas stays normal to the ionisation front (i.e. it does not respond to tangential pressure gradients), (ii) that the flow is time-independent, and (iii) that gravitational forces are negligible. 

%%%%%
\subsection{Radial flow from a cylindrical ionisation front}
%%%%%

We treat the flow of ionised gas close to the {\it Main Ionisation Front} as if the ionisation front were  semi-cylindrical. For purely radial flow off a cylindrical ionisation front, we can introduce a new variable, $\;r,\;$ which is the distance from the centre of curvature of the nearest part of the ionisation front. The ionisation front is at $\;r\!=\!Z_{_{\rm O}},\;$ so the equation of continuity becomes
\begin{eqnarray}\label{EQN:2Dcont_0}
r\,n(r)\,v(r)&=&Z_{_{\rm O}}\,n_{_{\rm II}}(5/3)^{1/2}c_{_{\rm II}}\;\,\equiv\;\,A_{_{\rm CYL}}^{1/2}\,,
\end{eqnarray}
where $A_{_{\rm CYL}}$ is defined here purely for mathematical convenience. The equation of motion then becomes
\begin{eqnarray}\nonumber
\frac{2}{A_{_{\rm CYL}}}\frac{dv}{dt}&=&\frac{2v}{A_{_{\rm CYL}}}\frac{dv}{dr}\;\,=\;\,\frac{d(v^2/A_{_{\rm CYL}})}{dr}\;\,=\;\,\frac{d}{dr}\left\{\frac{1}{r^2n^2}\right\}\\\label{EQN:2Dmotion_0}
&=&-\,\left\{\frac{2}{r^3n^2}+\frac{2}{r^2n^3}\frac{dn}{dr}\right\}\;\,=\;\,-\,\frac{2c_{_{\rm II}}^2}{A_{_{\rm CYL}}n}\frac{dn}{dr}\,.
\end{eqnarray}
If we now define dimensionless density, $\;\chi_{_{\rm CYL}}=n/n_{_{\rm II}},\;$ dimensionless velocity, $\;\nu_{_{\rm CYL}}= v/(5/3)^{1/2}c_{_{\rm II}},\;$  and dimensionless position, $\;\xi=r/Z_{_{\rm O}}\,$, this equation reduces to
\begin{eqnarray}\label{EQN:2Dden_0}
\frac{d\chi_{_{\rm CYL}}}{d\xi}&=&-\,\frac{\chi_{_{\rm CYL}}(\xi)}{\xi\left\{1 - 3\xi^2\chi_{_{\rm CYL}}^2(\xi)/5\right\}}\,,
\end{eqnarray}
and must be solved numerically, with boundary condition
\begin{eqnarray}\label{EQN:2DBC_0}
\chi_{_{\rm CYL}}(\xi\!=\!1)&=&1\,.
\end{eqnarray}
From Eqn. (\ref{EQN:2Dcont_0}), the dimensionless velocity is given by
\begin{eqnarray}\label{EQN:2Dvel_0}
\nu_{_{\rm CYL}}(\xi)&=&\frac{1}{\xi\chi_{_{\rm CYL}}(\xi)}\,.
\end{eqnarray}
Using Eqns. (\ref{EQN:2Dden_0}) to (\ref{EQN:2Dvel_0}) we obtain a universal solution for the dimensionless density and velocity in a cylindrical radial flow, as functions of the dimensionless radius. This solution can be scaled to any values of $\;n_{_{\rm II}},\;$ $\;c_{_{\rm II}}\;$ and $\;Z_{_{\rm O}}\;$ that we chose to adopt.

%%%%%
\subsection{Radial flow from a spherical ionisation front}
%%%%%

We treat the flow of ionised gas at large distances from the {\it Main Ionisation Front} as if the ionisation front were spherical. For purely radial flow off a spherical ionisation front of radius $Z_{_{\rm O}}$, the equation of continuity becomes
\begin{eqnarray}\label{EQN:3Dcont_0}
r^2\,n(r)\,v(r)&=&Z_{_{\rm O}}^2\,n_{_{\rm II}}(5/3)^{1/2}c_{_{\rm II}}\;\,=\;\,A_{_{\rm SPH}}^{1/2}\,,
\end{eqnarray}
where $A_{_{\rm SPH}}$ is again defined purely for mathematical convenience. The equation of motion then becomes
\begin{eqnarray}\nonumber
\frac{2}{A_{_{\rm SPH}}}\frac{dv}{dt}&=&\frac{2v}{A_{_{\rm SPH}}}\frac{dv}{dr}\;\,=\;\,\frac{d\left(v^2/A_{_{\rm SPH}}\right)}{dr}\;\,=\;\,\frac{d}{dr}\left\{\frac{1}{r^4n^2}\right\}\\\label{3Dmotion_0}
&=&-\left\{\frac{4}{r^5n^2}+\frac{2}{r^4n^3}\frac{dn}{dr}\right\}\;\,=\;\,-\frac{2c_{_{\rm II}}^2}{A_{_{\rm SPH}}n}\frac{dn}{dr}\,.
\end{eqnarray}
If we again define dimensionless density, $\;\chi_{_{\rm SPH}}=n/n_{_{\rm II}}\;$, dimensionless velocity, $\nu_{_{\rm SPH}}=v/(5/3)^{1/2}c_{_{\rm II}}$\,,  and dimensionless position, $\;\xi=r/Z_{_{\rm O}}\;$, this equation reduces to
\begin{eqnarray}\label{EQN:3Dden_0}
\frac{d\chi_{_{\rm SPH}}}{d\xi}&=&-\,\frac{2\chi_{_{\rm SPH}}(\xi)}{\xi\left(1-3\xi^4\chi_{_{\rm SPH}}^2(\xi)/5\right)}\,,
\end{eqnarray}
and must be solved numerically, with initial condition
\begin{eqnarray}\label{EQN:3DBC_0}
\chi_{_{\rm SPH}}(\xi =1)&=&1\,.
\end{eqnarray}
From Eqn. (\ref{EQN:3Dcont_0}), the dimensionless velocity is given by
\begin{eqnarray}\label{EQN:3Dvel_0}
\nu_{_{\rm SPH}}(\xi)&=&\frac{1}{\xi^2\chi_{_{\rm SPH}}(\xi)}\,.
\end{eqnarray}
Using Eqns. (\ref{EQN:3Dden_0}) to (\ref{EQN:3Dvel_0}) we obtain a universal solution for the dimensionless density and velocity in a spherical radial flow, as functions of the dimensionless radius. This solution can be scaled to any values of $\;n_{_{\rm II}},\;$ $\;c_{_{\rm II}}\;$ and $\;Z_{_{\rm O}}\;$ that we chose to adopt.

%%%%%
\subsection{The net flow of ionised gas}
%%%%%

In order to merge these two solutions, for the flow close to and far from the Main Ionisation Front, we 
introduce a factor
\begin{eqnarray}
\phi&=&1-{\rm e}^{-\xi/\zeta_\star}\,,
\end{eqnarray}
with $\zeta_\star = W_{_{\rm O}}/Z_{_{\rm O}}$. For $\xi\ll\zeta_\star$ (i.e. $r\ll W_{_{\rm O}}$), $\phi\ll 1$; and for $\xi\gg\zeta_\star$ (i.e. $r\gg W_{_{\rm O}}$), $\phi\sim 1$. Then Eqns. (\ref{EQN:2Dden_0}) and (\ref{EQN:3Dden_0}) can be combined to give
\begin{eqnarray}
\frac{d\chi_{_{\rm TOT}}}{d\xi}&=&-\,\frac{(1+\phi)\chi_{_{\rm TOT}}(\xi)}{\xi\left(1-3\xi^2(1+\phi\xi^2)\chi_{_{\rm TOT}}^2(\xi)/5\right)}\,, \\
\nu_{_{\rm TOT}}(\xi)&=&\frac{1}{\xi\,(1+\phi\,\xi)}
\end{eqnarray}
In this way the approximately radial/cylindrical flow near the ionisation front (i.e. $\xi\ll \zeta_\star$ or $r\ll Z_{_{\rm O}}$) merges smoothly into the approximately radial/spherical flow far from the star (i.e. $\xi\gg \zeta_\star$ or $r\gg Z_{_{\rm O}}$). The density in the flow is now given by $n=n_{_{\rm II}}\chi_{_{\rm TOT}}$, and the velocity by $v=(5/3)^{1/2}c_{_{\rm II}}\nu_{_{\rm TOT}}$.

Each point, $(x,y,z)$, in the H{\sc ii} region, is intercepted by the flows off two points on the {\it Main Ionisation Front}. We obtain the total density by summing contributions from these two flows, and the net velocity by summing their momenta. We then have the density and velocity fields throughout the H{\sc ii} region, viz.
\begin{eqnarray}
w&=&\left(x^2+y^2\right)^{1/2}\,,\\
r_{_1}&=&\left\{(W_{_{\rm O}}-w)^2+z^2\right\}^{1/2}\,,\\
r_{_2}&=&\left\{(W_{_{\rm O}}+w)^2+z^2\right\}^{1/2}\,,\\
\hat{\boldsymbol e}_{_1}&=&r_{_1}^{-1}\,\left\{\frac{x(W_{_{\rm O}}-w)}{w},\,\frac{y(W_{_{\rm O}}-w)}{w},\,z\right\}\,,\\
\hat{\boldsymbol e}_{_2}&=&r_{_2}^{-1}\,\left\{\frac{x(W_{_{\rm O}}+w)}{w},\,\frac{y(W_{_{\rm O}}+w)}{w},\,z\right\}\,,\\
n(x,y,z)&=&n_{_{\rm II}}\,\left\{\chi_{_{\rm TOT}}\!\left(\!\frac{r_{_1}}{W_{_{\rm O}}}\!\right)+\chi_{_{\rm TOT}}\!\left(\!\frac{r_{_2}}{W_{_{\rm O}}}\!\right)\right\}\,,\\\nonumber
v(x,y,z)&=&\left(\!\frac{5}{3}\!\right)^{\!1/2}c_{_{\rm II}}\!\left\{\chi_{_{\rm TOT}}\!\left(\!\frac{r_{_1}}{W_{_{\rm O}}}\!\right)+\chi_{_{\rm TOT}}\!\left(\!\frac{r_{_2}}{W_{_{\rm O}}}\!\right)\right\}^{-1}\\\nonumber
&&\hspace{0.5cm}\times\left\{\chi_{_{\rm TOT}}\!\left(\!\frac{r_{_1}}{W_{_{\rm O}}}\!\right)\nu_{_{\rm TOT}}\!\left(\!\frac{r_{_1}}{W_{_{\rm O}}}\!\right)\hat{\boldsymbol e}_{_1}\right.\\
&&\hspace{1.0cm}\left.+\;\chi_{_{\rm TOT}}\!\left(\!\frac{r_{_2}}{W_{_{\rm O}}}\!\right)\nu_{_{\rm TOT}}\!\left(\!\frac{r_{_2}}{W_{_{\rm O}}}\!\right)\hat{\boldsymbol e}_{_2}\right\}.
\end{eqnarray}

Although these formulae for merging the two regimes (near to, and far from, the ionisation front) are rather arbitrary, we argue that this is unlikely to corrupt the results significantly, since the emission measure is dominated by the ionised gas close to the ionisation front, where the flow must approximate well to radial/cylindrical symmetry.

%%%%%
%%%%%
\begin{figure}
\begin{center}
\hspace{3.5cm}\includegraphics[width=8cm,angle=270]{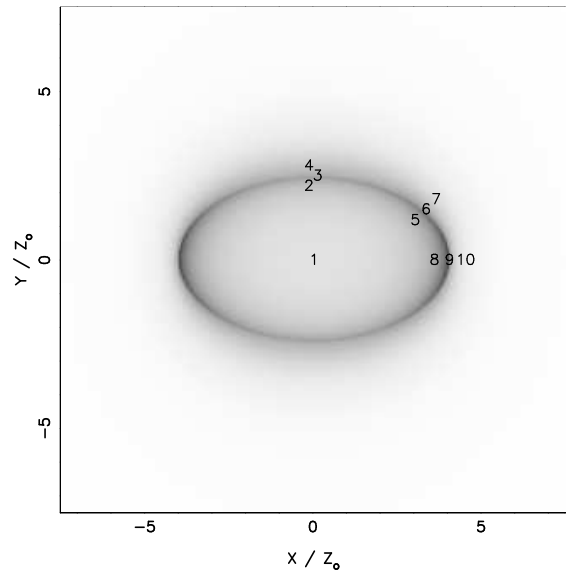} 
\end{center}
\caption{Optically-thin freefree emission-measure map for a configuration with $W_{_{\rm O}}=5Z_{_{\rm O}}$, viewed at angle $\theta =46.6^{\rm o}$ to the axis of symmetry. The units are arbitrary, and the grey-scale is linear. The co-ordinates of the frame are scaled to $Z_{_{\rm O}}$. The numbers mark the lines of sight along which the recombination-line profiles illustrated in Fig. \ref{FIG:46.6_RRLineProfiles} are obtained.}
\label{FIG:46.6_FreeFreeMap}
\end{figure}
%%%%%

%%%%%
\begin{figure}
 \begin{center}
 \hspace{1.0cm}\includegraphics[width=8cm,angle=270]{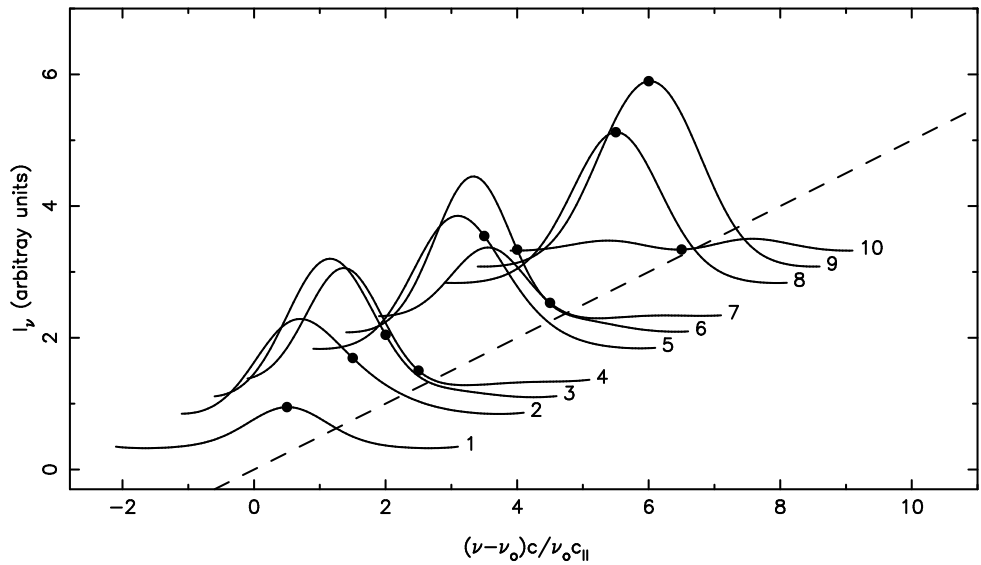} 
 \end{center}
\caption{Recombination line profiles from the lines of sight marked on Fig. \ref{FIG:46.6_FreeFreeMap}. The profiles are displaced parallel to the dashed line, to reduce confusion, and the zero-velocity point on each profile is marked with a filled circle.}
\label{FIG:46.6_RRLineProfiles}
\end{figure}
%%%%%

%%%%%
\begin{figure}
 \begin{center}
\hspace{2.0cm}\includegraphics[width=8cm,angle=270]{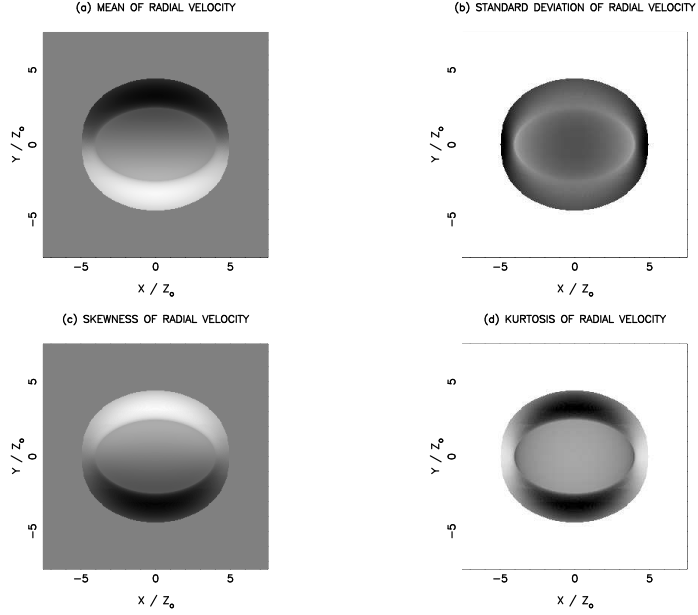} 
 \end{center}
\caption{Maps of (a) the mean, (b) the standard deviation, (c) the skewness and (d) the kurtosis of the radial velocity of recombination lines emitted from a configuration with $W_{_{\rm O}}=5Z_{_{\rm O}}$, viewed at angle $\theta =46.6^{\rm o}$ to the axis of symmetry. All grey-scales are linear. The mean and standard deviation are normalised to $(5/3)^{1/2}c_{_{\rm II}}$, and the corresponding grey-scale ranges are, respectively, $(-1.0,+1.0)$ and $(0.0,1.5)$. The skewness and kurtosis are dimensionless and the corresponding grey-scale ranges are, respectively, $(-7.0,+7.0)$ and $(0.0,20.0)$. }
\label{FIG:46.6_RRLineMoments}
\end{figure}
%%%%%
%%%%%

%%%%%
\section{The transport of continuum and line radiation} \label{SEC:RadTrans}
%%%%%

To generate intensity maps and line profiles for this configuration, we define a suite of viewing angles, $\theta$, relative to the symmetry axis ($\hat{\bf k}$); the direction to the observer is then defined by a unit vector
\begin{eqnarray}
\hat{\boldsymbol e}_{_{\rm OBS}}&=&\left\{\sin(\theta),0,\cos(\theta)\right\}\,,
\end{eqnarray}
so positions ${\boldsymbol R}\equiv(X,Y)$ on the plane of the sky through the ionising star correspond to positions in the configuration frame, $(x,y,z)$, given by
\begin{eqnarray}
x&=&-\cos(\theta)\,Y\,,\\
y&=&X\,,\\
z&=&\sin(\theta)\,Y\,.
\end{eqnarray}

To construct a map of the freefree emission, in the optically thin limit, we divide this plane into square pixels, and compute the emission measure through the centre of each pixel, ${\boldsymbol R}_{_p}\equiv(X_{_p},Y_{_p})$, i.e.
\begin{eqnarray}\label{EQN:EmMeas}
{\cal EM}_{_p}&=&\int\limits_{s=-4W_{_{\rm O}}}^{s=+4W_{_{\rm O}}}\;n^2\!\left({\boldsymbol R}_{_p}+\hat{\boldsymbol e}_{_{\rm OBS}}\,s\right)\,ds
\end{eqnarray}
It is straightforward to scale this integral for a particular system, so as to give the freefree intensity, at any wavelength where the freefree emission is optically thin. For a particular system it is also straightforward to adjust the integral so that it allows for the freefree emission being optically thick at low frequencies.

To construct the recombination line profile at a particular pixel, $p$, we assume the configuration has zero systemic velocity, and define a suite of discrete radial velocities, $v_{_q}$, regularly spaced with separation $\Delta v$, in the range $-4c_{_{\rm II}}\leq v_{_q}\leq +4c_{_{\rm II}}$. For a line having rest frequency $\nu_{_{\rm O}}$, the intensity at the corresponding frequency $\nu_{_q}=\nu_{_{\rm O}}(1-v_{_q}/c)$, is given by
\begin{eqnarray}\label{EQN:RRLProf}
I_{_{p.q}}&\propto&\int\limits_{s=-4W_{_{\rm O}}}^{s=+4W_{_{\rm O}}}\;n^2\!\left({\boldsymbol R}_{_p}+\hat{\boldsymbol e}_{_{\rm OBS}}\,s\right)\;\frac{c}{\nu_{_{\rm O}}}\;\psi\!\left(v_{_q}\right)\;ds\,.
\end{eqnarray}
Here, the constant of proportionality depends on the physical parameters of the configuration {\it and} the particular recombination line being considered. $\;\psi(v)$ is the 1D distribution of random microscopic velocities. For the purpose of illustration, we assume a thermal distribution, so
\begin{eqnarray}
\psi(v)&=&\frac{1}{(2\pi)^{1/2}\,c_{_{\rm II}}}\exp\!\left\{\frac{-\,\left(v-\left[{\boldsymbol v}\!\left({\boldsymbol R}_{_p}+\hat{\boldsymbol e}_{_{\rm OBS}}\,s\right)\cdot\hat{\boldsymbol e}\right]\right)^2}{2\,c_{_{\rm II}}^2}\right\}.
\end{eqnarray}

To construct maps of the mean radial velocity, its standard deviation, skewness and kurtosis, we need, for each pixel, the first four moments of the radial velocity along the corresponding line of sight, viz.
\begin{eqnarray}
{\cal M}_{_{p.n}}&\simeq&\sum\limits_{q}^{}\,\left\{I_{_{p.q}}\,v_{_q}^n\right\}\;\Delta v\,.
\end{eqnarray}
It is then straightforward to compute the mean radial velocity, its standard deviation, skewness and kurtosis.

%%%%%
\begin{figure}
\begin{center}
\hspace{3.5cm}\includegraphics[width=8cm,angle=270]{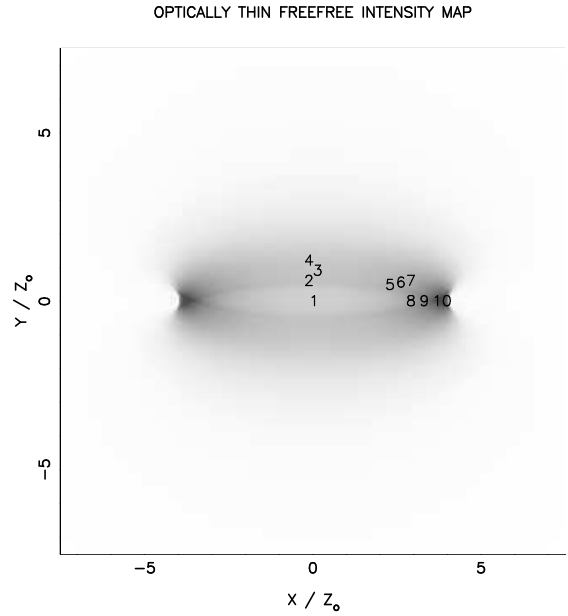} 
\end{center}
\caption{As Fig. \ref{FIG:46.6_FreeFreeMap}, but viewed at an angle $\theta =71.8^{\rm o}$ to the line of sight, and the numbers now mark the lines of sight along which the recombination-line profiles illustrated in Fig. \ref{FIG:71.8_RRLineProfiles} are obtained.}
\label{FIG:71.8_FreeFreeMap}
\end{figure}
%%%%%

%%%%%
\begin{figure}
 \begin{center}
\hspace{1.0cm}\includegraphics[width=8cm,angle=270]{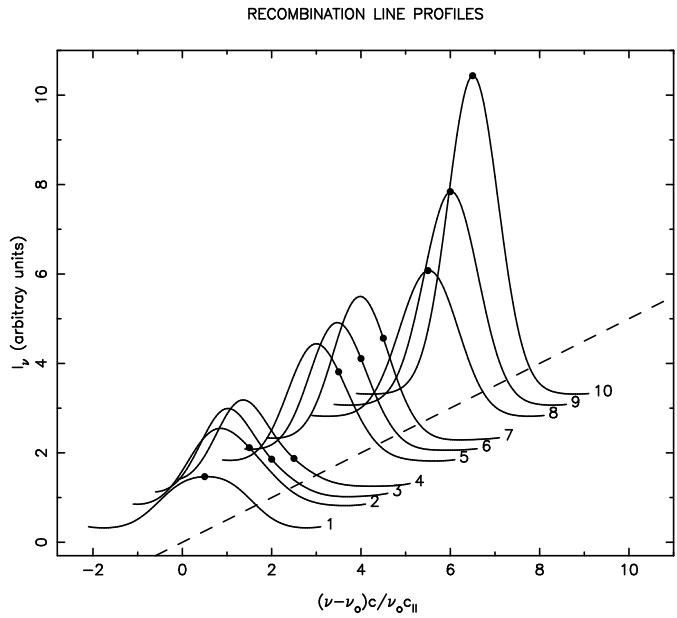} 
 \end{center}
\caption{As Fig. \ref{FIG:46.6_RRLineProfiles}, but for a viewing angle $\theta =71.8^{\rm o}$ from the axis of symmetry, and from the lines of sight marked on Fig. \ref{FIG:71.8_FreeFreeMap}.}
\label{FIG:71.8_RRLineProfiles}
\end{figure}
%%%%%

%%%%%
\begin{figure}
 \begin{center}
\hspace{2.0cm}\includegraphics[width=8cm,angle=270]{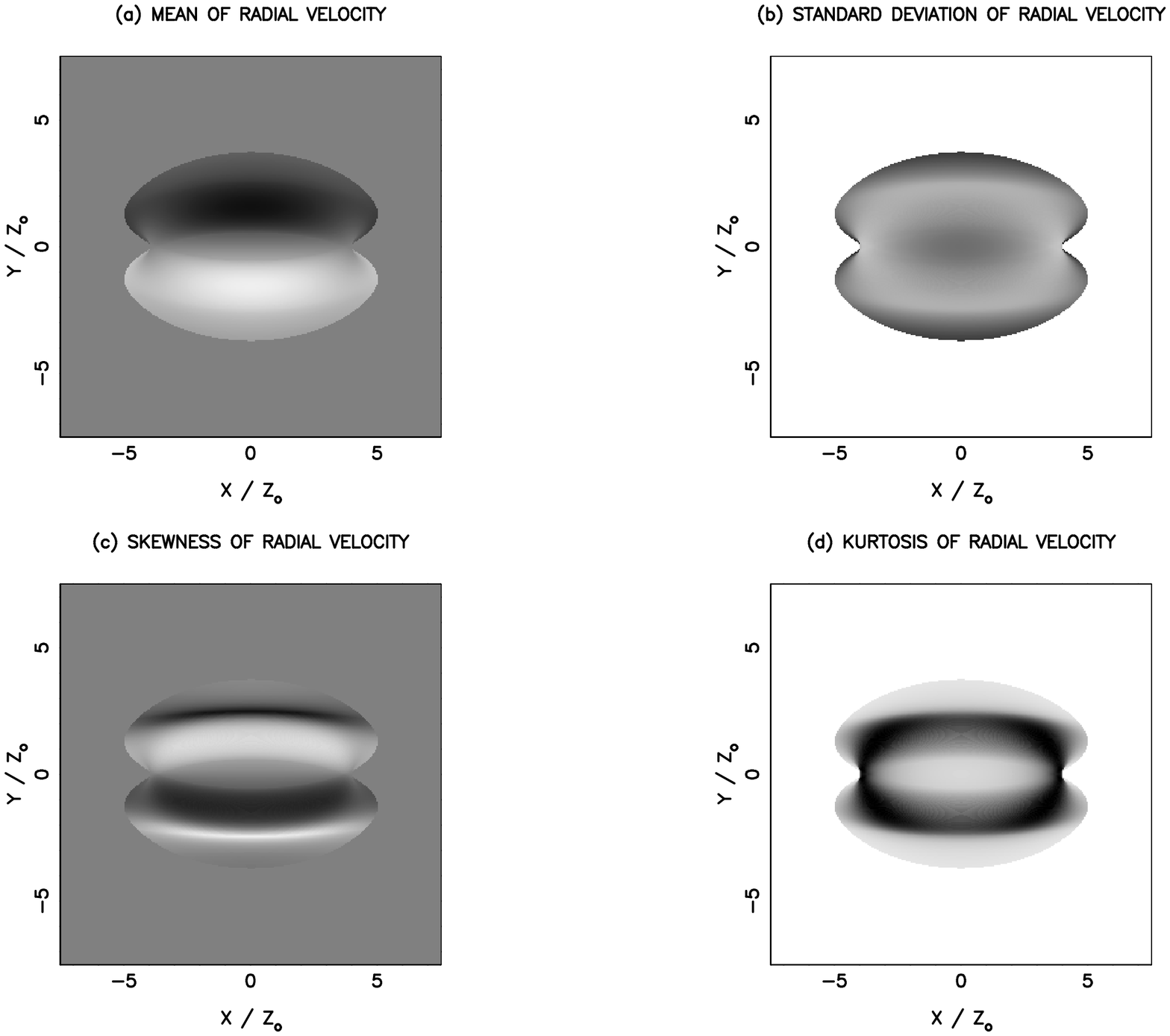} 
 \end{center}
\caption{As Fig. \ref{FIG:46.6_RRLineMoments}, but from lines of sight at angle $\theta =71.8^{\rm o}$ to the axis of symmetry.}
\label{FIG:71.8_RRLineMoments}
\end{figure}
%%%%%

%%%%%
\section{Results}\label{SEC:Results}
%%%%%

We present results for a configuration with $\zeta_\star=W_{_{\rm O}}/Z_{_{\rm O}}=5$. This means that the ratio between the distance from the star to the nearest points on the ionisation front and the thickness of the layer is
\begin{eqnarray}
\frac{(W_{_{\rm O}}-Z_{_{\rm O}})}{2Z_{_{\rm O}}}&=&2\,,
\end{eqnarray}
and we note that -- due to the assumption of optical thinness -- this ratio is the only parameter that we need to specify. We consider two representative viewing angles, $\theta = 46.6^{\rm o}$ (closer to the pole) and $\theta = 71.8^{\rm o}$ (further from the pole). 

Fig. \ref{FIG:46.6_FreeFreeMap} shows the emission measure map (Eqn. \ref{EQN:EmMeas}) from viewing angle $\theta = 46.6^{\rm o}$. At high frequencies where the freefree emission is optically thin, this map can also be read as representing the freefree emission. The dominance of the dense gas boiling off the ionisation front is clear. We note that the drop in the gas density away from the ionisation front is not simply due to the divergence of the flow, but also due to the acceleration of the material as it flows down the density gradient. 

Fig. \ref{FIG:46.6_RRLineProfiles} shows the line profiles (Eqn. \ref{EQN:RRLProf}) at the representative points marked with numbers 1 to 10 on the emission measure map shown in Fig. \ref{FIG:46.6_FreeFreeMap}. The profiles are displaced relative to one another to reduce confusion, and the dashed line marks the direction of the displacement; the solid circle on each profile marks the point at zero velocity. 

Although some of the line profiles illustrated on Fig. \ref{FIG:46.6_RRLineProfiles} are distinctive, in a real source with marked departures from the symmetries assumed here it would be hard to look for these. We have therefore computed maps characterising the statistical properties of the radial velocity of recombination lines,  on the principle that these may contain more robust signature of the velocity field characterising our model. These are presented in Fig. \ref{FIG:46.6_RRLineMoments}, which gives -- reading from left to right and top to bottom -- maps of (a) the mean, (b) the standard deviation, (c) the skewness and (d) the kurtosis of the radial velocity of the ionised gas.

The same information is given for the same  configuration ($W_{_{\rm O}}/Z_{_{\rm O}}=5$) but viewed from an angle $\theta =71.8^{\rm o}$, further from the axis of symmetry, in Figs. \ref{FIG:71.8_FreeFreeMap}, \ref{FIG:71.8_RRLineProfiles} and \ref{FIG:71.8_RRLineMoments}. Again the optically thin freefree emission map highlights the ring-like concentration of ionised gas at the waist of the bipolar H{\sc ii} region, and the line profiles are in places distinctive, but it would seem that the radial velocity statistics (mean, standard deviation, skewness and kurtosis) offer the best chance of establishing the flow pattern in the ionised gas.

%%%%%
\section{Conclusions}\label{SEC:Conc}
%%%%%

We have developed a model for the formation of a bipolar H{\sc ii} region, following a collision between two clouds. The collision is presumed to lead to the formation of a shock compressed layer, and a star cluster then condenses out near the centre of the layer. At the same time the layer contracts laterally, feeding additional material towards the cluster. Provided the cluster includes at least one ionising star that stays near the mid-plane of the layer, or more than one star with at least one on either side of the mid-plane, the H{\sc} region excited by the stars expands rapidly in the directions orthogonal to the layer, to produce a bipolar H{\sc ii} region; in the plane of the layer expansion of the H{\sc ii} regions stalls where it meets the ram pressure  of the in-falling gas from the outer reaches of the layer. 

With a view to testing this hypothesis, we have developed the computational machinery required to generate maps of freefree continuum emission and recombination-line profiles, from configurations characterised by the net ionising output of the star cluster, $\dot{\cal N}_{_{\rm LyC}}$, the thickness of the layer, $2Z_{_{\rm O}}$, and the radius of the ionised hole in the layer, $W_{_{\rm O}}\!-\!Z_{_{\rm O}}$. 

We illustrate this capability for a representative configuration, in which, provided we assume the emission is optically thin, we only need to specify $W_{_{\rm O}}/Z_{_{\rm O}}=5$. We show that freefree maps are useful to trace the waist of the bipolar H{\sc ii} region, and we suggest that it may be easiest to analyse the dynamics of the ionised gas by mapping the mean, standard deviation skewness and kurtosis of the radial velocities of radio recombination lines. 

We are planning an observational programme to look for these signatures in known bipolar H{\sc ii} regions.

%%%%%
\begin{figure}
 \begin{center}
 \end{center}
\caption{This is the first figure.}\label{fig:sample}
\end{figure}
%%%%%

%%%%%
\begin{ack}
APW gratefully acknowledges the support of a consolidated grant (ST/K00926/1), from the UK Science and Technology Funding Council, and thanks the Laboratoire d'Astrophysique de Marseille for their hospitality when this project was started . This work was performed using the computational facilities of the Advanced Research Computing at Cardiff (ARCCA) Division, Cardiff University. All false-colour images have been rendered with SPLASH \citep{PriceD2007}.
\end{ack}
%%%%%

%%%%%

%%%%%

\end{document}